# Advanced Testing Chain Supporting the Validation of Smart Grid Systems and Technologies

Ron Brandl*, Panos Kotsampopoulos†, Georg Lauss‡, Marios Maniatopoulos†, Maria Nuschke*,
Juan Montoya*, Thomas I. Strasser‡, Diana Strauss-Mincu*

*Fraunhofer Institute for Energy Economics and Energy System Technology, Kassel, Germany,
email: {ron.brandl, maria.nuschke, juan.montoya, diana.strauss-mincu}@iee.fraunhofer.de
†NTUA National Technical University of Athens, Athens, Greece, email: {kotsa@power.ece.ntua.gr, mmaniato@mail.ntua.gr}
‡AIT Austrian Institute of Technology, Vienna, Austria, email: {georg.lauss, thomas.strasser}@ait.ac.at

*Abstract*—New testing and development procedures and methods are needed to address topics like power system stability, operation and control in the context of grid integration of rapidly developing smart grid technologies. In this context, individual testing of units and components has to be reconsidered and appropriate testing procedures and methods need to be described and implemented. This paper addresses these needs by proposing a holistic and enhanced testing methodology that integrates simulation/software- and hardware-based testing infrastructure. This approach presents the advantage of a testing environment, which is very close to field testing, includes the grid dynamic behavior feedback and is risks-free for the power system, for the equipment under test and for the personnel executing the tests. Furthermore, this paper gives an overview of successful implementation of the proposed testing approach within different testing infrastructure available at the premises of different research institutes in Europe.

## I. Introduction

Modern power systems are characterized by the integration of distributed-renewable generation posing new challenges but at the same time opening new possibilities. The increasing complexity of the power system requires the transition into a smart grid highlighting the importance of Information and Communication Technology (ICT) [1], [2]. Therefore, intelligent control strategies are required to manage such networks. The effective laboratory-based testing and validation of such new control strategies is an important step prior to field deployment [3]. Nowadays advancement in real-time computation served to a significant increase of power system testing solutions. Methods and solutions like pure software simulation, coupled simulation (co-simulation), Hardware-in-the-Loop (HIL) experiments and pure hardware-based testing as depicted Fig. 1 enhance todays testing methods [4], [5].

The advantage of the testing chain proposed in this paper, compared to pure simulation studies of a developed controller in a holistic approach (integrating power system models, components, cyber-physical systems, grid protection schemes, ICT, etc.), consists in adding flexibility and realism to the test scenarios by, for instance, replacing simulation models with real hardware. Furthermore, co-simulation platforms integrate controller schemes, weather data and real on-line measurement data into a real-time simulation environment providing the possibility to perform investigations that are more realistic. Next step after the execution of pure software simulation studies is Controller Hardware-in-the-Loop (CHIL) and Power Hardware-in-the-Loop (PHIL) investigations [4], [5], [6], which enable the testing of hardware controller and physical power devices connected to a real-time simulation. In the end laboratory and field test are mostly required for finalizing the validation [3], [7].

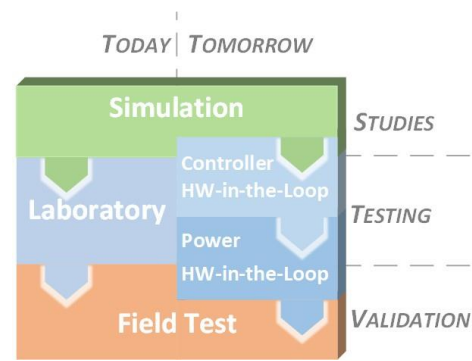

Fig. 1. Overview of current and proposed testing and validation methods.

The remaining part of this paper is structured as follows: Section II outlines necessary requirements for future power system testing whereas Section III discuss the evolution of HIL-based testing methods for electric power systems. The concept of an advanced testing chain is proposed in Section IV followed by a proof-of-concept validation using selected examples in Section V. Finally, the paper is summarized and concluded with Section VI.

## II. Requirements and Needs of Future Testing

To understand the needs of a general testing chain approach for smart grid controls, the basic controller model developed in the framework of the European Horizon 2020 project ERIGrid is presented in Fig. 2 [8].

It can be seen that this model integrates different controller levels (i.e., D1-D5) and communication interfaces (i.e., L1-L4). A Microgrid Controller (MGC) for instance is located in the control level D3 – "Intelligent Electronic Device (IED)". As shown in this concept, the MGC will not control the DER unit (D5) directly, the unit is managed by the DER controller represented in level D4. Although the D4 level is, in current practice, together with D5 and their communication (L4 and L3) in general simplified to only represent the main behavior of the DER, but the complete operation and interfaces of the MGC controlled units are essential and needs to be tested in a holistic way.

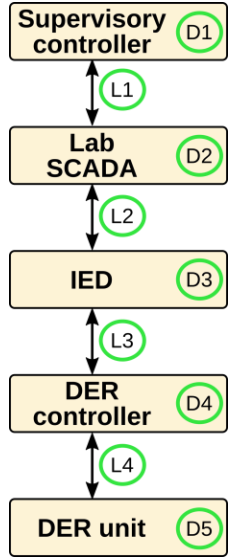

Fig. 2. Basic controller model (modified from [8]).

HIL-based setups enable testing according to the above introduced concept. However, to study and validate the entire functionality and behavior of new power system scenarios, all kind of preliminary system studies, control algorithm testing and device validation need to go alongside with each other [3]. In this paper a generic way of a holistic testing chain for developing new power system enhancements in the frame of smart grid technologies is introduced. To do so, HIL methods are used to provide a more detailed and realistic behavior during the hardware testing phase and to serve a chain link between studies, testing and validation.

## III. EVOLUTION OF HARDWARE-IN-THE-LOOP METHODS FOR ELECTRIC POWER SYSTEMS

In recent years, HIL-based approaches for investigations on electric power systems attract more and more interest from the power system community because of the given advantages compared to classical methodologies such as pure laboratory testing or numerical simulations [5], [6]. Related to this real-time based HIL simulation domain, two basic approaches can be differentiated and identified with the PHIL and the CHIL method, respectively [9].

### A. Power Hardware-in-the-Loop Method

The PHIL simulation represents an advanced methodology for system analysis, testing, and validation. In this context it is exclusively used for electrical power systems, and already considered as an important technology for testing smart grid systems and components such as Distributed Energy Resource (DER)/Distributed Generator (DG) devices, energy storage systems, and Distributed Flexible AC Transmission System (DFACTS) [9], [10]. PHIL simulation features the connection of an actual power system defined as the Hardware-under-Test (HUT) to a real-life system defined as the Rest-of-System (ROS). The ROS is usually simulated in a Digital Real-Time Simulator (DRTS), allowing repeatable and economical testing under realistic, highly flexible and scalable conditions [4], [5], [9]. Using this PHIL setup, extreme conditions can be tested with reduced risks, while deviations of normal operation modes can be emulated and identified for both the ROS and the HUT. Therefore, an in-depth understanding of the tested device, its steady state and transient system behavior can be revealed.

PHIL testing combines the advantages of numeric simulation and physical hardware testing [4]. Because of its hybrid origin, the PHIL approach is particularly suitable for studying the integration of new components such as Photovoltaic (PV) generation plants, wind turbines, electric vehicles, or energy storage systems in power supply networks. Even entire Microgrids can be connected to a simulated active network, containing various simulated DER devices, where complex interactions can be identified [4], [5], [6]. Since now, research related to PHIL simulation focused primarily on single components testing for the purpose of characterization and validation. A move towards integrated system-level testing by means of power and ICT sub-systems applying the PHIL simulation method is necessary for future applications.

### B. Controller Hardware-in-the-Loop Method

The CHIL approach represents a simulation technique, which can be seen as connatural to the PHIL approach. It is of major interest for electric power systems engineers because of reasons of flexibility and capabilities for accelerated rapid prototyping. The peripherals of an electromagnetic system are simulated in a real-time software environment that is directly connected to a physical control board which represents the HUT. Instead of having a real power hardware component linked with a simulated power system defined as ROS, only the controller of the power hardware component is available as physical device defined as the new HUT.

Compared to PHIL simulation, no high-power amplification unit is existent between software and the hardware systems, but only low-level signal amplifiers are used for the CHIL simulation. CHIL simulation are mainly executed in real-time, however, also non-real-time simulations are possible depending on the operation and the implementation of the controller [4], [5], [9].

### C. Software-in-the-Loop Method

It should be mentioned at this stage, that the Software-in-the-loop (SIL) simulation represents a further methodology. It is characterized by the fact that both the electric power system defined as the ROS and the physical device defined as the HUT are executed on the same PC or DRTS hardware. It is widely used in combination with off-line computing and software validation of digital controllers for electric power systems, in general [11].

### D. Combination of Methods

From a system perspective, pure off-line digital simulations using mathematical models have been used for the emulation of power system phenomena in the past. However, the large-scale deployment of complex devices representing the state-of-the-art, such as power electronics-based DER, offer novel control functionalities and constitute new challenges for simulation. The above-mentioned devices are particularly difficult to model in an accurate way and might be involved in complex interactions within the electric power system. Therefore, testing by means of PHIL and CHIL simulation can reveal system related phenomena that are not visible in pure digital simulations [1], [4], [6], [9], [12].

Fig. 3 depicts a testing and validation platform combining several domains suitable for the execution of CHIL and PHIL simulation and presents a possible combination of applying both methodologies for power system testing. As shown in Fig. 3, the Virtual Simulated System (VSS), executing power system models on a DRTS, serves as a link between the Component Control System (CCS) containing, inter alia, Controller, Protection, and the Physical Power System (PPS) where power units are located. Several other domains like ICT systems can easily be added [13].

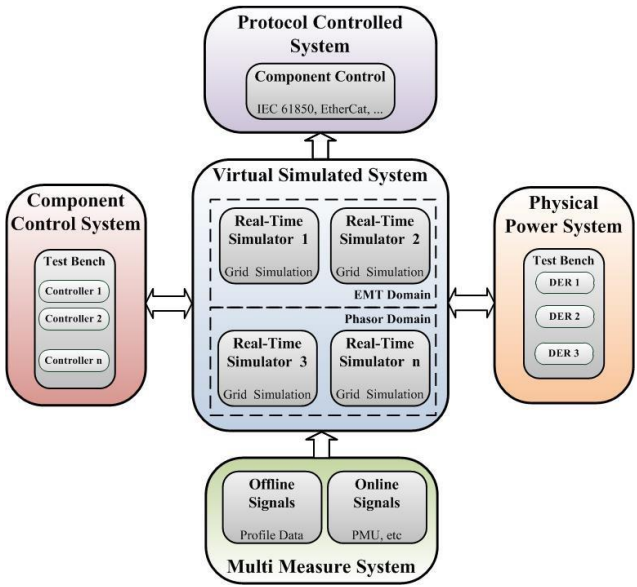

Fig. 3. Platform for CHIL and PHIL simulation for real-time based investigations of electric power systems [13].

## IV. CONCEPT OF AN ADVANCED TESTING APPROACH

As outlined above an advanced approach for testing and validating smart grid systems, concepts, and corresponding technologies is required. In the following the concept of such a testing chain is introduced and discussed covering mainly four different stages.

### A. Testing Chain Approach for Smart Grids

Fig. 4 provides an overview of the proposed testing chain. In *Stage 1* investigations performed in a pure software simulated environment are usually carried out. This means, that the electric power system as well as the so-called "one-to-be-tested" control algorithm is part of the VSS. The power system is simulated in steady state or transient conditions. It is of benefit to use simulation tools that already exist and which can be easily adapted to the later test cases in order to perform real-time simulations of the power system model. Typically, the "to-be-tested" algorithm is located in and executed by the same simulation tool, respectively VSS as the power system model. This enables the functionality test of the control algorithm, but does not represent adequately the interface between power and control systems (e.g., L3 and L4 of the basic controller model are neglected).

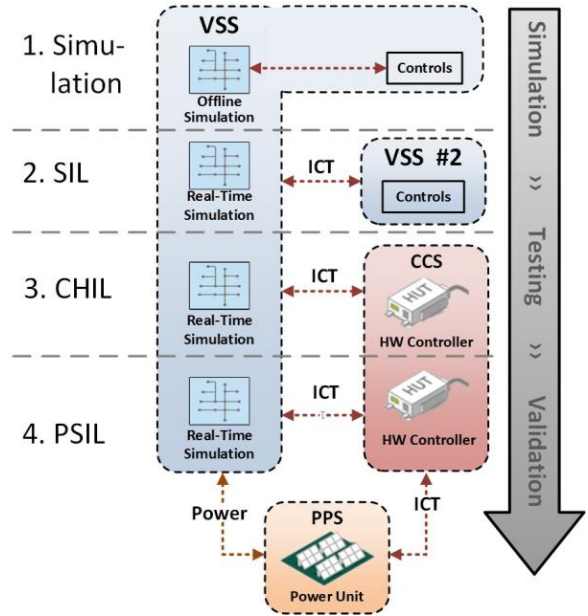

Fig. 4. Testing chain approach for smart grids systems [14].

Proceeding the development of the defined control algorithm, *Stage 2* of the testing chain propose the use of two dedicated software tools for executing the power system model and controller separately. This SIL simulation or co-simulation technique allows the exchange of information in a closed-loop configuration. This means that data between one VSS domain will be continuously sent to the other and a response will be fed back to the origin. In general, both VSS domains can be either located in the same simulation tool with an additional simulated communication emulator or can be placed on different simulators connected via a digital interface to exchange information.

After verifying the correct behavior of the control algorithm in Stage 1 and 2, *Stage 3* deals especially with the performance validation of the actual hardware controller by the use of a CHIL setup. To perform a CHIL experiment, the power system model needs to be executed in real-time on a DRTS where the control algorithm is implemented on the hardware controller. This can be a level D4 power component controller such as DER controllers, or relays. It can also include a level D3 centralized controller such as MGC, virtual power plant controllers, distribution management system controller, or a Web-of-Cell controller [15] according to the categorization highlighted in Fig. 2.

CHIL testing provides significant benefits compared to simulation-only and SIL experiments. Using DRTS for executing power system models in real-time, the actual hardware controller can be tested including all kinds of communication interfaces and potential analog signal measurements by interfacing it with the DRTS. Moreover, the DRTS can perform realistic power system conditions on the one hand but furthermore can generate extreme conditions for performance studies of the hardware controller. Thus, up-to-now hidden weaknesses of the control algorithm can be studied such as time delays, noise, signal probe inaccuracy, processor performance issues, etc. However, since all power components are still simulated by the DRTS, a CHIL test only represents the interconnection of the controller with ideal physical power devices, not taking into account additional inaccuracies, physical hardware limitations and communication issues.

The final *Stage 4*, before actual field-testing and implementation, of the proposed testing chain approach is the integration of real physical power hardware controlled by the hardware controller. This combined CHIL and PHIL is called Power System-in-the-Loop (PSIL) and includes the controller as well as power apparatus like inverter, motors, etc . This technique offers the closest possibility of a field test, which can be implemented in a laboratory, since it integrated real-time interactions between the hardware controller, the physical power component and the simulated power system test case executed on the DRTS. In spite of the high complexity to ensure stable, safe and accurate experiments, a PSIL setup enables an investigation, not as a single and separate entity, but as a holistic power system. This technique is proofed to validate entire functionalities of real hardware controller, interdependencies and interactions between real power components in an entire flexible and repetitive laboratory environment.

### B. Benefits and Challenges

One major challenge of the proposed testing chain is the high complexity of implementing realistic interfaces between the different domains (e.g., VSS, PPS, and CCS). Each step to the next stage of the testing chain will increase the complexity of the interface, but increase the realistic behavior of the entire system step-by-step and therefore is mandatory to verify the developed power system enhancement. As beginning with simulation-only test performance, no interface issues are present since all connections are neglected because the entire system is implemented in the same VSS.

Executing SIL tests (i.e., Stage 2), a wide variation of interfaces possibility is offered. Different communication interfaces can be implemented and tested according to best use (SunSpec/Modbus, CAN, IEC 61850, etc.). The simulation of actual field communication during this initial state of development de-risks issues, which normally will appear during pure hardware or field test implementation. Thus, validating the software compatibility and synchronize data exchange between the systems is of high interest to limit cost and time resources of a new power system development.

Heading to Stage 3 of the testing chain, CHIL investigation provides a huge advantage by adding realistic condition and real-life performance of the hardware controller. By interfacing hardware devices in real-time through digital and analog signals, several mismatches and realistic challenges can be taken into account:
- Signal and measurement inaccuracy,
- Voltage operating limits of analog / digital port,
- Noise and time delays of transmitted signals, and
- Communication characteristics.

Furthermore, by integrating a communication emulator additional studies can be made, like time delay, bandwidth, packet loss and other cyber-physical aspects [16]. Contrary to CHIL setups, in PHIL simulation non-realist considerable interface challenges appear [17], [18], [19]. Mainly the mismatch between operating voltages are of high importance. Low-level signals of DRTS contrary to power signals of the power apparatus requires additional voltages adaption and power production / consumption. This requires a power amplifier, which integrates additional undesired disturbance to the system together with measurement inaccuracy of probes that are necessary to feed back the behavior of the HUT. It is most important to guarantee the stability of the PHIL system, since it can lead to damage laboratory equipment. At several cases achieving stability may reduce the accuracy of the experimental results. Therefore, knowledge of the used components and expected issues are mandatory when performing a PHIL experiment. To perform an entire PSIL investigation, a wide range of possibility to test and validate smart grid system and technology enhancements are offered.

## V. Proof-of-Concept Validation

Two examples are provided and discussed below in order to present the use of the above proposed testing chain approach.

### A. Development of a Off-grid System Micro Grid Controller

The combination of CHIL and PHIL simulation allows to test a controller in a CHIL configuration, while critical power components are incorporated as physical devices (PHIL simulation) allowing to analyze more realistic real-world based conditions. For example, a MGC is developed to implement secondary control in the operation of off-grid systems. An island system is considered which consists of a diesel generator, PV generation, loads and lines. The main task of the MGC is to ensure that the minimum load ratio (30% of nominal power) of the diesel generator is respected at times of high irradiation. Respecting the minimum load ratio of the diesel generator is important in order to reduce tear and wear and maintenance needs. The MGC measures the active power of the loads and the PV production and, based on its algorithm, will send activation commands to the load banks.

The MGC is tested in pure simulation, SIL, CHIL and finally combined CHIL and PHIL, according to the proposed testing chain of Fig. 4. The power network dynamic model is executed on the RTDS® with a time-step of 50 $\mu$s. At the first stages of the testing chain the load banks were purely simulated, therefore an ideal controllable behavior was considered. However, at the last stage (PSIL experiment) limitations of the physical load banks had to be considered (available steps of power, delays, etc.), showing the benefit of the combined CHIL/PHIL approach.

The laboratory setup of the PSIL test is shown in Fig. 5. The activation of the load bank is controlled by the laboratory supervisory control running on Labview software. Currently a similar concept is developed, where instead of the use of load banks, the PV inverter curtails its active power according to set-points provided by the MGC. In this way additional equipment is not necessary (i.e., load banks), however a recently acquired PV inverter should be used in order to support this functionality.

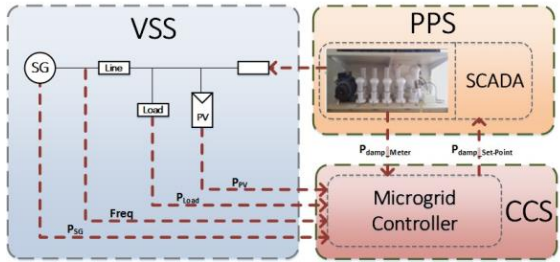

Fig. 5. Laboratory set-up for testing MGCs for off-grid systems at ICCS-NTUA, Greece

### B. Development of a Blackstart capable Microgrid Controller

Another MGC has been developed in order to fulfill several tasks for the purpose of investigations of Microgrids at medium voltage level:

- Initiation, coordination and monitoring of restoration process of a Microgrid after blackout (loss of mains),
- Energy management during islanding operation,
- Adaptive protection setting control, and
- Resynchronization and connection to the main grid after main voltage recovery.

The investigated Microgrid consists of a grid-forming PV-battery (VSI) inverter, a second battery inverter with a conventional current controlled (CSI) approach and an aggregated load. During the development and validation of this MGC the proposed development chain has been utilized. After modeling of all physical components of the Microgrid as well as development of the MGC algorithms in simulation environment the operation of the MGC has been tested in pure simulation and SIL setup in order to validate the general behavior and functionality (see Fig. 6, upper left case). This is equivalent to Stage 1 and 2 of the proposed testing chain. Next step was the compilation of the control algorithm of the MGC for the target hardware, which was a Programmable Logic Controller (PLC).

In the CHIL stage the PLC was connected to the Microgrid model running on a DRTS in order to verify its technical performance and communication (see Fig. 6, upper right case). To perform a PHIL simulation experiment, real power apparatus replaced parts of the simulated Microgrid. In this fourth stage of the proposed testing chain, the PLC is managing parts of the simulated Microgrid and sends set-points to the real battery inverter, which is connected to the DRTS via a power amplifier. The experiment validated communication between the PLC and inverter as well as the MCG performance with the real power hardware during the several test case performances of the DRTS (see Fig. 6, lower left case). The final step in the testing will be the integration of the MGC into a pure hardware setup to validate the system in a field test experiment (see Fig. 6, lower left case).

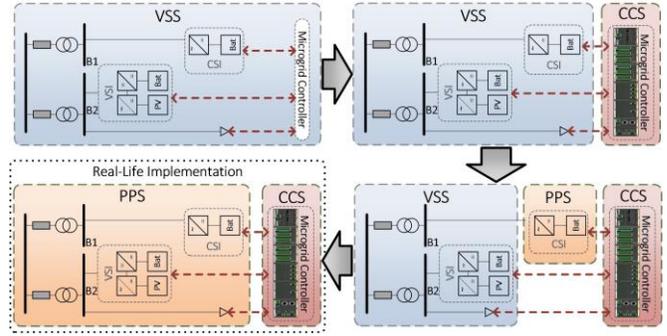

Fig. 6. Testing Process for a Controller Development at Fraunhofer IEE, Germany [20].

## VI. CONCLUSIONS

Pure software simulation, SIL, co-simulation, CHIL and PHIL, as well as field test implementations are nowadays techniques to support the development of new smart grid technologies. The holistic testing methodology proposed in this paper is easy to implement and parts of the chain can be reused in different testing stages (e.g. the simulation model in the VSS domain and the hardware controller in the CCS domain can be used at several stages). Consequently, existing test results can serve as reference for new tests and the systematical approach is beneficial to development processes in the following points:

- Systematical analysis during all stages of a development,
- Early detection of functionality, performance, communication and interdependency issue,
- Time and cost saving, due to lower effort needed in field test implementation.

The testing chain approach presented in this paper is already in use and available within several research infrastructure. The same applies for the methodology to analyze step-by-step new Smart Grid technologies in a holistic manner. The proposed general and efficient presented methodology is of advantage not only for research, but also for industry due to its time and cost efficient practice. Future works will mainly concentrate on the refinement of the proposed testing chain as well as its application to additional, more complex scenarios.

## ACKNOWLEDGMENT

This work is partly supported by the European Communitys Horizon 2020 Program (H2020/2014-2020) under project

"ERIGrid" (Grant Agreement No. 654113). The participation of AIT within IEA ISGAN/SIRFN is funded in the frame of the IEA Research Cooperation program by the Austrian Ministry for Transport, Innovation and Technology (Contract No. FFG 839566).